\documentclass[sigconf]{acmart}
\usepackage{graphicx}
\usepackage{subcaption}
\usepackage[utf8x]{inputenc}
\usepackage{xcolor}
\usepackage{gensymb}
\usepackage{booktabs}
\AtBeginDocument{%
  \providecommand\BibTeX{{%
    \normalfont B\kern-0.5em{\scshape i\kern-0.25em b}\kern-0.8em\TeX}}}

\copyrightyear{2019}
\acmYear{2019}
\setcopyright{rightsretained}

\acmConference{CASCON'19}{November 4-6, 2019}{Toronto, ON, Canada}
\acmDOI{10.1145/3306307.3328180}
\acmISBN{978-1-4503-6317-4/19/07}
\acmBooktitle{CASCON'19, November 4-6, 2019, Toronto, ON, Canada}
\begin{document}

\title{Classification of Histopathological Biopsy Images Using Ensemble of Deep Learning Networks}

\author{Sara Hosseinzadeh Kassani}
\email{sara.kassani@usask.ca}
\orcid{0000-0002-5776-7929}

\affiliation{%
  \institution{University of Saskatchewan}
  \city{Saskatoon}
  \country{Canada}
}

\author{Peyman Hosseinzadeh Kassani}
\email{peymanhk@tulane.edu}
\affiliation{%
  \institution{University of Tulane}
  \city{New Orleans}
  \country{USA}}

\author{Michal J. Wesolowski}
\email{mike.wesolowski@usask.ca}
\affiliation{%
  \institution{University of Saskatchewan}
  \city{Saskatoon}
  \country{Canada}
}

\author{Kevin A. Schneider}
\email{kevin.schneider@usask.ca}
\affiliation{%
  \institution{University of Saskatchewan}
  \city{Saskatoon}
  \country{Canada}
}

\author{Ralph Deters}
\email{deters@cs.usask.ca}
\affiliation{%
 \institution{University of Saskatchewan}
 \city{Saskatoon}
 \country{Canada}}






\begin{abstract}
    Breast cancer is one of the leading causes of death across the world in women. Early diagnosis of this type of cancer is critical for treatment and patient care. Computer-aided detection (CAD) systems using convolutional neural networks (CNN) could assist in the classification of abnormalities. In this study, we proposed an ensemble deep learning-based approach for automatic binary classification of breast histology images. The proposed ensemble model adapts three pre-trained CNNs, namely VGG19, MobileNet, and DenseNet. The ensemble model is used for the feature representation and extraction steps. The extracted features are then fed into a multi-layer perceptron classifier to carry out the classification task. Various pre-processing and CNN tuning techniques such as stain-normalization, data augmentation, hyperparameter tuning, and fine-tuning are used to train the model. The proposed method is validated on four publicly available benchmark datasets, i.e., ICIAR, BreakHis, PatchCamelyon, and Bioimaging. The proposed multi-model ensemble method obtains better predictions than single classifiers and machine learning algorithms with accuracies of 98.13\%, 95.00\%, 94.64\% and 83.10\% for BreakHis, ICIAR, PatchCamelyon and Bioimaging datasets, respectively. 
\end{abstract}

\ccsdesc[500]{Computing methodologies~Artificial intelligence}
\ccsdesc[500]{Computing methodologies~Object recognition}
\ccsdesc[500]{Computing methodologies~Machine learning approaches}
\ccsdesc[300]{Computing methodologies~Supervised learning by classification}

\keywords{Computer-aided diagnosis, Deep learning, Feature extraction, Multi-model ensemble, Transfer learning}
\maketitle
\section{Introduction}
    Breast cancer has become one of the major causes of cancer-related death worldwide in women~\cite{KHAN20191}. According to the World Health Organization reports~\cite{WHO_report}, in 2018, it is estimated that 627,000 women died from invasive breast cancer - that is approximately 15\% of all cancer-related deaths among women and breast cancer rates are increasing in nearly every country globally. It is evident that early detection and diagnosis plays an essential role in effective treatment planning and patient care. Cancer screening using breast tissue biopsies aims to distinguish between  benign or malignant lesions. However, manual assessment of large-scale histopathological images is a challenging task due to the variations in appearance, heterogeneous structure, and textures~\cite{LI2019165}. Such a manual analysis is laborious, and time intensive and often dependent on subjective human interpretation. For this reason, developing CAD systems is a possible solution for classification of Hematoxylin-Eosin (H\&E) stained histological breast cancer images. In recent years, deep learning outperformed state-of-the-art methods in various fields of machine learning and medical image analysis tasks, such as classification~\cite{mardanisamani2019crop},  detection~\cite{herent2019detection}, segmentation~\cite{LATEEF2019321}, and computer-based diagnosis~\cite{MAIER201986}. The merit of deep learning compared to other types of learners is its ability to obtain the performance similar to or better than human performance. Feature extraction is a critical step since the classifier performance directly depends on the quality of extracted low and high-level features. Several feature fusion methods employing pre-trained CNN models were proposed in the literature that effectively applied to medical imaging applications~\cite{PERDOMO2019181, MA2019143, AMINNAJI2019201}. Motivated by the success of ensemble learning models in computer vision, we propose a novel multi-model ensemble method for binary classification of breast histopathological images. The experimental results on four publicly available datasets demonstrate that the proposed ensemble method generates more accurate cancer prediction than single classifiers and widely-used machine learning algorithms. 

\section{Related works}
    Developing CAD systems using digital image processing and deep learning algorithms can assist pathologists with better diagnostic accuracy and less computational time. In~\cite{VO2019123}, a combination of CNN and the boosting trees classifier was proposed for breast cancer detection on BreakHis dataset. The proposed model employed Inception-ResNet-v2 model for visual feature extraction from multi-scale images. Then a boosting classifier using gradient boosting trees was used for final classification step. In~\cite{pratiher2019diving}, an ensemble of histological hashing and class-specific manifold learning was proposed for both binary and multi-class breast cancer detection on BreakHis dataset. In~\cite{ROY201990}, a patch-based classifier by CNN and majority voting method were used for breast cancer histopathology classification on the augmented ICIAR dataset. The proposed classifier predicts the class label on both binary and multi-class task. In~\cite{GANDOMKAR201814}, a framework using deep residual network was developed for H\&E histopathological image classification. In~\cite{han2017breast}, a deep learning method based on GoogLeNet architecture was used for the image classification task, and a majority voting method was used for patient-level classification. In~\cite{bejnordi2017context}, a context-aware stacked convolutional neural network architecture was used for classifying whole slide images. The proposed method was trained on large input patches extracted from tissue structures. Finally, in~\cite{spanhol2016breast}, a deep learning method based on AlexNet architecture was used to classify breast histopathological images as benign or malignant cases.

\begin{figure}[h]

  \centering
  \includegraphics[width=\linewidth]{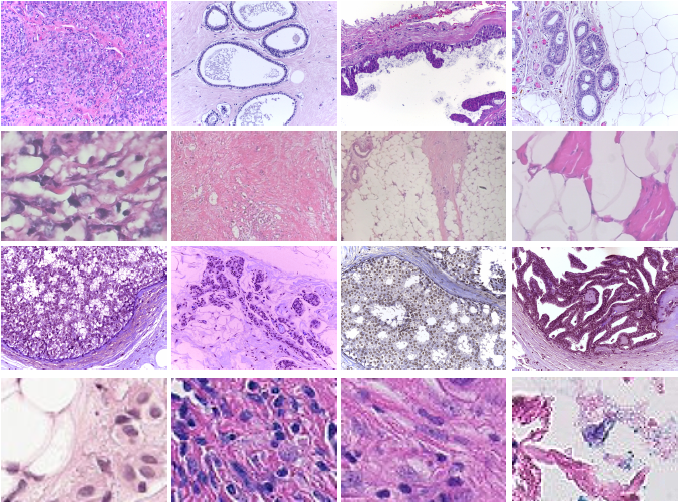}
  \caption{Examples of variability in tissue patterns. Bioimaging 2015 (first row),
   BreakHis (second row), ICIAR 2018 (third row) and, PatchCamelyon dataset (fourth row).}
   \label{fig:differentDatasets}
\end{figure}

    A number of visual characteristics such as variations in sources of acquisition device, different protocols in stain normalization, variations in color, and heterogeneous textures in histopathological slide images can affect the performance of the Deep CNNs~\cite{li2019weakly}. Hence, developing a robust automated analysis tool to support the issue of data heterogeneity collected from multiple sources is a major challenge. To address this challenge, we propose a novel three-path ensemble architecture for binary classification of breast histopathological images collected from different datasets. Figure \ref{fig:differentDatasets} depicts some examples of histology images acquired from different datasets. The variability and similarity of provided datasets can be observed in this figure. 

    The main contribution of this work is proposing a generic method that does not need handcrafted features and can be easily adapted to different datasets with the aim of reducing the generalization error and obtaining a more accurate prediction. We compared obtained results with the traditional machine learning algorithms and also with each selected CNN individually. Experimental results showed that the proposed method outperforms both the state-of-the-art architectures and the traditional machine learning algorithms on the provided datasets. The proposed model employs three well-established pre-trained CNNs - VGG19, MobileNet, and DenseNet which aims to incorporate specific components, i.e., standard convolutions, separable convolutions, depthwise convolutions, long skip, and short-cut connections. Doing so, we are able to overcome the data heterogeneity constraint and efficiently extract discriminative image features. 
    
    The rest of this paper is organized as follows. The proposed methodology for automatically classifying benign and malignant tissues is explained in Section 3. The datasets' description, experimental settings, hyperparameter optimization and performance metrics are given in Section 4. A brief discussion and results analysis are provided in Section 5, and finally, the conclusion is presented in Section 6.

\section{Methodology}
\subsection{Proposed Network architecture}
\begin{figure}[h]
  \centering
  \includegraphics[width=0.6\linewidth]{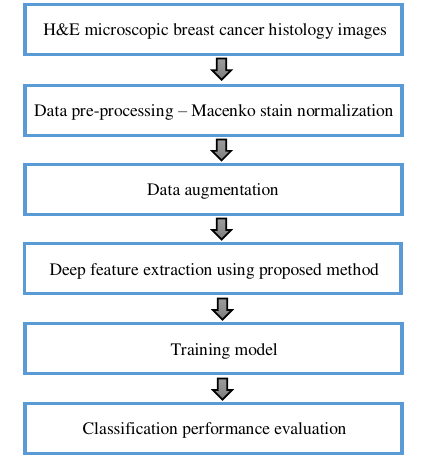}

  \caption{Block diagram of the proposed methodology.}
  \label{fig:blockDiagram}
\end{figure}

    Few studies have been published on the application of the ensemble deep learning method to breast histopathology images. Each of the adapted CNN architectures in the proposed model are constructed by different types of convolution layers in order to promote feature extraction and aggregation of fundamental information from a given input image. The block diagram of the proposed methodology of this study is shown in Figure~\ref{fig:blockDiagram}. As it can be seen in this figure, the entire methodology is mainly divided into six steps: collecting H\&E microscopic breast cancer histology images, data pre-processing, data augmentation, feature extraction using the proposed network, classification and finally model evaluation. We first improved the quality of visual information of each input image using different pre-processing strategies. Then the training dataset size is increased with various data augmentation techniques. Once input images are prepared, they are fed into the feature extraction phase with the proposed ensemble architecture. The extracted features from each architecture are flattened together to create the final multi-view feature vector. The generated feature vector is fed into a multi-layer perceptron to classify each image into corresponding classes. Finally, the performance of the proposed method is evaluated on test images using the trained model. We validated the performance of our proposed CNN architecture on the four publicly available datasets, namely: ICIAR, BreakHis, PatchCamelyon and Bioimaging. 

\subsection{Feature extraction using transfer learning}
    Considering the high visual complexity of histopathological images, proper feature extraction is essential because of its impact on the performance of the classifier. However, due to the privacy issue in the medical domain~\cite{uchibeke2018blockchain}, the provided datasets are not large enough to sufficiently train a CNN~\cite{hu2018deep}. Recently, blockchain technology has been foreseen as a solution in the area of healthcare for secure data ownership management of electronic medical data or medical IoT devices~\cite{samaniego2018smart,samaniego2019pushing}. Aiming to tackle this challenge, a transfer learning strategy has been widely investigated to exploit the knowledge learned from cross domains instead of training a model from scratch with randomly initialized weights. In this method, we transfer knowledge learned by a dataset into the new dataset in another domain. Using a transfer learning approach, the model can learn general features from a source dataset that do not exist in the current dataset. Transfer learning has advantages such as speeding up the convergence of the network, reducing the computational power, and optimizing the network performance~\cite{lu2019pathological}. 

\subsection{Three-path ensemble architecture for breast cancer classification}
\begin{figure}[h]
  \centering
  \includegraphics[width=0.8\linewidth]{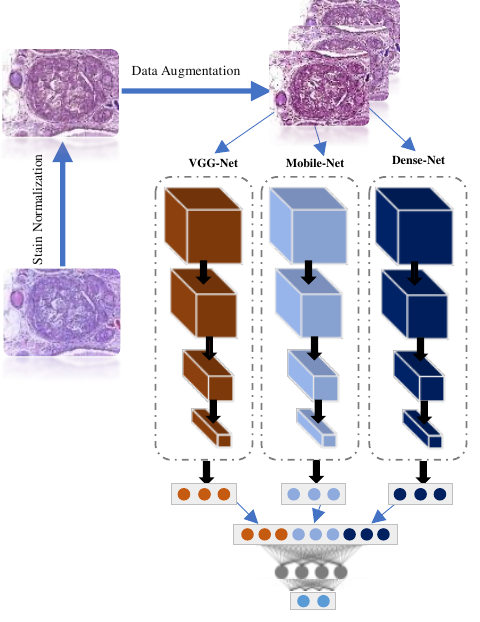}
  \caption{The proposed ensemble network with a three-path CNN of VGGNet, MobileNet and DenseNet.}
  \label{fig:proposedArchitecture}
\end{figure}

    Three well-known architectures, VGG19~\cite{simonyan2014very}, MobileNetV2~\cite{howard2017mobilenets} and DenseNet201~\cite{huang2017densely} are selected based on their (i) satisfying performances in different computer vision tasks (ii) usefulness towards real-time (or near real-time) applications and, (iii) feasibility of transfer learning for limited datasets. Considering that each method has shortcomings in regards to the variations of the shape and texture of the input image, inspired by the work of~\cite{moeskops2016automatic}, we propose a three-path ensemble prediction approach to make use of the advantages of the multiple classifiers to improve overall accuracy. We selected theses networks based on the obtained results of an exhaustive grid-search technique on different state-of-the-art architectures (i.e. InceptionV3, InceptionresNetV2, Xception, ResNet50, MobileNetV2 and DenseNet201, VGG19  and VGG16) with different combination of hyperparameters including, optimizer, learning rate, weight initialization, batch size, dropout rate to obtain the best possible performance for breast cancer detection.
    Figure~\ref{fig:proposedArchitecture} illustrates the proposed ensemble architecture for breast cancer classification. As demonstrated in Figure~\ref{fig:proposedArchitecture}, the proposed architecture is constructed by three independent CNN architectures. The final fully connected layers of each CNN architecture are combined together to produce the final feature vector. This combination allows capturing more informative features. Therefore, it is possible to achieve a more robust accuracy. 
    
    VGGNet~\cite{simonyan2014very} was introduced by Karen Simonyan and Andrew Zisserman from Visual Geometry Group (VGG) of the University of Oxford in 2014. It achieves one of the top performances in the ImageNet Large Scale Visual Recognition Challenge (ILSVRC) 2014. The network used 3$\times$3 convolutional layers stacked on top of each other, alternated with a max pooling layer, two 4096 nodes for fully-connected layers, and finally followed by a softmax classifier. 
    
    The MobileNet~\cite{howard2017mobilenets} architecture is the second model used for this study. MobileNet, designed by Google researchers, is mainly designed for mobile phones and embedded applications. The MobileNet architecture was built based on depth-wise separable convolutions, followed by a pointwise convolution with a 1$\times$1 convolution layer. In the standard convolution layer, each kernel is applied to all channels on the input image. While depthwise convolution is applied on each channel separately. This approach significantly reduces the number of parameters once is compared to standard convolutions with the same depth. MobileNet achieved inspiring performance over various applications with a fewer number of hyperparameters and computational resources. 
    
    As our third feature extractor, we employed DenseNet~\cite{huang2017densely}  architecture. DenseNet, stands for Densely-Connected Convolutional Networks, is proposed by Huang et al.~\cite{huang2017densely}. DenseNet introduces dense block, which is a sequential of convolutional layers, wherein every layer has a direct connection to all subsequent layers. This structure solves the issue of vanishing gradient and improves feature propagation by using very short connections between input and output layers throughout the network.

\section{Experiments}
\subsection{Datasets description}
    Four benchmark datasets are used for evaluating the performance of the proposed model. BreakHis~\cite{spanhol2016breast} dataset consisting of 7909 H\&E stained microscopic images which was collected from 82 anonymous patients. The dataset is divided into benign and malignant tumor biopsies. Small patches were extracted at four magnification of $\times$40, $\times$100, $\times$200, and $\times$400. The benign tumors were classified into four subclasses which were adenosis (A), tubular adenoma (TA), phyllodes tumor (PT), and fibroadenoma (F) and the malignant tumors were also classified into four subclasses which were ductal carcinoma (DC), mucinous carcinoma (MC), lobular carcinoma (LC), and papillary carcinoma (PC). 
    
    A modified version of the Patch Camelyon (PCam) benchmark dataset~\cite{veeling2018rotation, bejnordi2017diagnostic}, publicly available at~\cite{kagglehistodatase}, consisting of benign and malignant breast tumor biopsies is also used to evaluate the performance of the proposed classification model. The dataset consists of 327,680 microscopy images with 96$\times$ 96-pixel size patches extracted from the whole-slide images with a binary label indicating the presence of metastatic tissue. We used the modified version of this database since the original Patch Camelyon database contained duplicated images.
    
    Additionally, two other datasets, the Bioimaging 2015~\cite{bioimaging2015dataset} challenge dataset and the ICIAR 2018~\cite{aresta2019bach} dataset, are used in this work. The ICIAR 2018 dataset, available as part of the BACH challenge, was an extended version of the Bioimaging 2015 dataset. Both datasets consisted of 24 bits RGB H\&E stained breast histology images and extracted from whole slide image biopsies, with a pixel size of 0.42 $\mu$m $\times$ 0.42 $\mu$m acquired with 200$\times$ magnification. Each image is classified into four different classes, namely: normal tissues, benign lesions, in situ carcinomas and invasive carcinomas. The Bioimaging dataset contained 249 microscopy training images and 36 microscopy testing images in total, equally distributed among the four classes. The ICIAR dataset contained 100 images in each category, i.e., in a total of 400 training images. In order to create the binary database from these two datasets, we grouped the normal and benign classes into the benign category and the in situ and invasive classes into the malignant category.

\subsection{Data preparation and pre-processing techniques}
    We adopted different data preparation techniques such as data augmentation, stain-normalization and image normalization strategies to optimize the training process. In the following, we briefly explain each of them. 
\subsubsection{Data augmentation}
\begin{figure}[h]
  \centering
  \includegraphics[width=\linewidth]{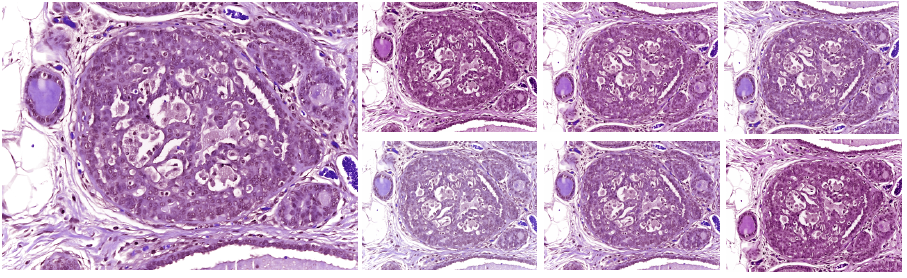}
  \caption{Images obtained after data augmentation techniques. The left image is the original image and the right images are the artificially generated image after different data augmentation methods}
  \label{fig:dataAugmentation}
\end{figure}

    Due to the limited size of the input samples, training the CNN is prone to over-fitting leading to low detection rate~\cite{li2019benign}. One solution to alleviate this issue is the data augmentation technique in which the aim is to generate more training data from the existing training set~\cite{kassani2019comparative}. Different data augmentation techniques, such as horizontal flipping, rotating and zooming are applied to datasets to create more training samples. The data augmentation parameters utilized for all datasets are presented in Table~\ref{tab:dataAugParameters}. Examples of histopathological images after the augmentation are shown in Figure~\ref{fig:dataAugmentation}.

\begin{table}[ht]
\centering
\renewcommand{\arraystretch}{1.2}
\caption{Data augmentation parameters.}
 \label{tab:dataAugParameters}
\begin{tabular}{ll}
\hline
\textbf{Parameter}   & \textbf{Value} \\ \hline
Horizontal Flip      & True           \\
Vertical Flip        & True           \\
Contrast Enhancement & True           \\
Zoom Range           & 0.2            \\
Shear Range          & 0.2            \\
Rotational Range     & 90$^{\circ}$             \\ 
Fill Mode            & Nearest        \\ \hline
\end{tabular}
\end{table}

\subsubsection{Stain-normalization}
     The tissue slices are stained by Haematoxylin and Eosin (H\&E) to differentiate between nuclei stained with purple color as well as other tissue structures stained with pink and red color to help pathologists analyze the shape of nuclei, density, variability and overall tissue structure. However, H\&E staining variability between acquired images exists due to the different staining protocols, scanners and raw materials which is a common problem with histological image analysis. Therefore, stain-normalization of  H\&E stained histology slides is a necessary step to reduce the color variation and obtain a better color consistency prior to feeding input images into the proposed architecture. Different approaches have been proposed for stain normalization in histological images including Macenko et al.~\cite{macenko2009method}, Reinhard et al.~\cite{reinhard2001color} and Vahadane et al.~\cite{vahadane2015structure}. For this experiment, Macenko et al. [22] approach is applied due to its promising performance in many studies~\cite{xu2018automated,ROY201990,albarqouni2016aggnet,saraswat2014automated} to standardize the color intensity of the tissue. Macenko method is based on a singular value decomposition (SVD). 
     In this method, a logarithmic function~\cite{macenko2009method} is used to adaptively transform color concentration of the original histopathological image into its optical density (OD) image as given in equation 1. 
 \begin{equation}
  OD = - log \left ( \frac{I}{I_{0}} \right )
\end{equation}
    Where OD is the matrix of optical density values, I is the image intensity in RGB space and $ I_{0} $ is the illuminating intensity incident on the histological sample.

\subsubsection{Image normalization}
    Another necessary pre-processing step is intensity normalization. The primary purpose of image normalization~\cite{Yu2017} is to obtain the same range of values for each input image before feeding to the CNN model which also helps to speed up the convergence of the model. Input images are normalized to the standard normal distribution by min-max normalization to the intensity range of [0, 1], which is computed as: 
 \begin{equation}
  x{_{norm}} = \frac{x-x_{min}}{x_{max} - x_{min}}
\end{equation}
    where X is the pixel intensity. $x_{min}$ and $x_{max}$ are minimum and maximum intensity values of the input image in equation 2.

\subsection{Experimental settings}
    All images were resized to 224x224 pixels using bicubic interpolation according to the input size of the selected pre-trained models. The batch size was set to 32 and all models trained for 1000 epochs. A fully connected layer trained with the rectified linear unit (ReLU) activation function with 256 hidden neurons followed by a dropout layer with a probability of 0.5 to prevent over-fitting. Dropout layer helps to further reduce over-fitting by randomly eliminates their contribution in the training process. For Adam optimizer, $\beta 1 $, $\beta 2$ and learning rate were set to 0.6, 0.8 and 0.0001, respectively. For fine-tuning, we have modified the last dense layer in all architectures to output two classes corresponding to benign and malignant lesions instead of 1000 classes as was proposed for ImageNet. All pre-trained Deep CNN models are fine-tuned separately. Also, the network weights were initialized from weights trained on ImageNet. The operating system is Windows with an Intel(R) Core(TM) i7-8700K 3.7 GHz processors with 32 GB RAM. Training and testing process of the proposed architecture for this experiment is implemented in Python using Keras package with Tensorflow as the deep learning framework backend and run on Nvidia GeForce GTX 1080 Ti GPU with 11GB RAM.

\subsection{Evaluation criteria}
    The performance of the proposed classification model evaluated based on recall, precision, F1-score, and accuracy. Given the number of true positives (TP), false positives (FP), true negatives (TN) and false negatives (FN), the measures are mathematically expressed as follows: 
 \begin{equation}
    Accuracy = \frac{TP+TN}{TP+TN+FP+FN}\times 100 
 \end{equation}

 \begin{equation}
   Precision = \frac{TP}{TP+FP}\times 100
\end{equation}

 \begin{equation}
   Recall = \frac{TP}{TP+FN}\times 100
\end{equation}

 \begin{equation}
   F1-Score=2 \times \frac{Recall \times Precision}{Recall + Precision}
\end{equation}

\section{Discussion }
\begin{figure}[h]
  \centering
  \includegraphics[width=\linewidth]{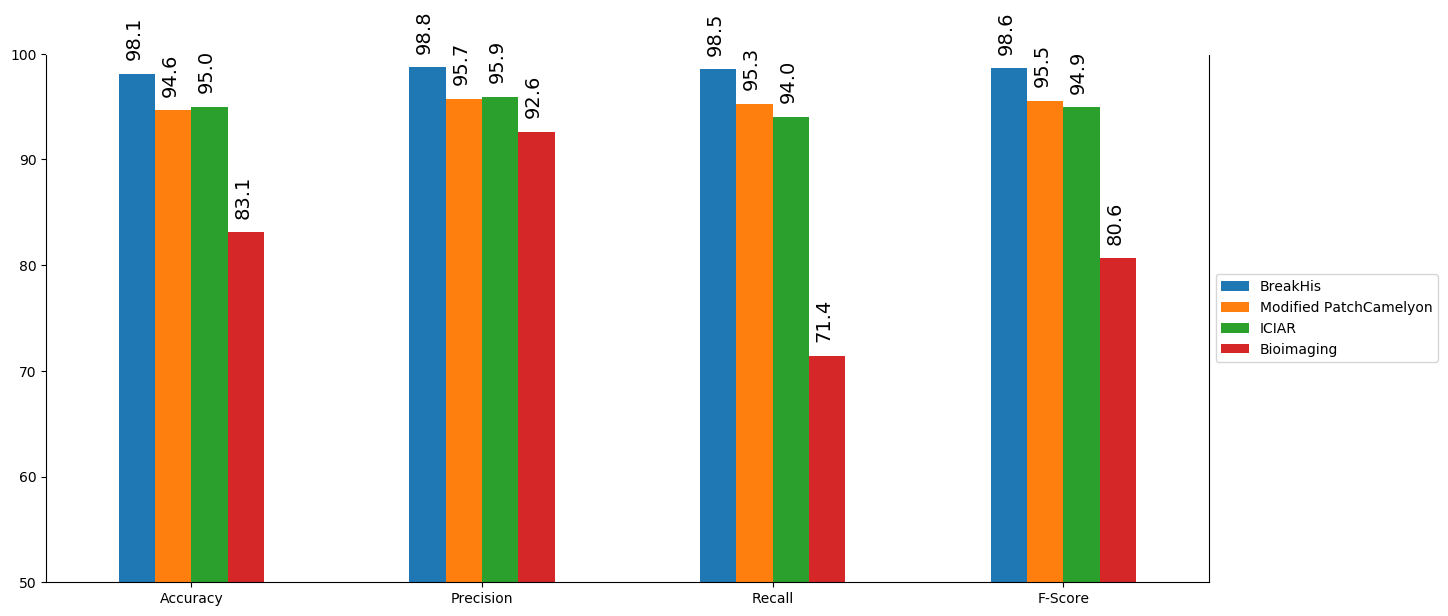}
  \caption{Results of accuracy, precision, recall, and F-score of the proposed method on four open access datasets}
  \label{fig:barchartEnsembleresults}
\end{figure}

    In this research, we focused on the binary classification for histopathological images using a three-path ensemble architecture with transfer learning and fine-tuning. To verify the effectiveness of the presented methodology, different comparative analyses were conducted. First, we compare the obtained results of the proposed ensemble model on the four provided datasets. Then, the comparison between proposed ensemble architecture and CNN classifiers individually is provided and finally, we present the comparison of the proposed ensemble architecture and machine learning algorithms.
    In Table~\ref{tab:ensembleresults} and Figure~\ref{fig:barchartEnsembleresults}, the obtained accuracy, precision, recall and F-score of the proposed approach for each benchmark dataset is demonstrated. The proposed method on BreakHis dataset achieved the highest accuracy, precision, recall, and F-score with values of 98.13\%, 98.75\%, 98.54\% and 98.64\%, respectively. 
    \begin{table}[h]
    \caption{Results of accuracy, precision, recall, and F-score of the proposed method on four open access datasets.}
    \label{tab:ensembleresults}
    \begin{tabular}{lllll}
    \toprule
                   & Accuracy & Precision & Recall  & F-score \\
    \midrule
    BreakHis       & 98.13\%  & 98.75\%   & 98.54\% & 98.64\% \\
    PatchCamelyon* & 94.64\%  & 95.70\%   & 95.27\% & 95.50\% \\
    ICIAR          & 95.00\%  & 95.91\%   & 94.00\% & 94.94\% \\
    Bioimaging     & 83.10\%  & 92.60\%   & 71.42\% & 80.64\%\\
  \bottomrule
 
\end{tabular}
\end{table}

\begin{figure}[h]
  \centering
  \includegraphics[width=\linewidth]{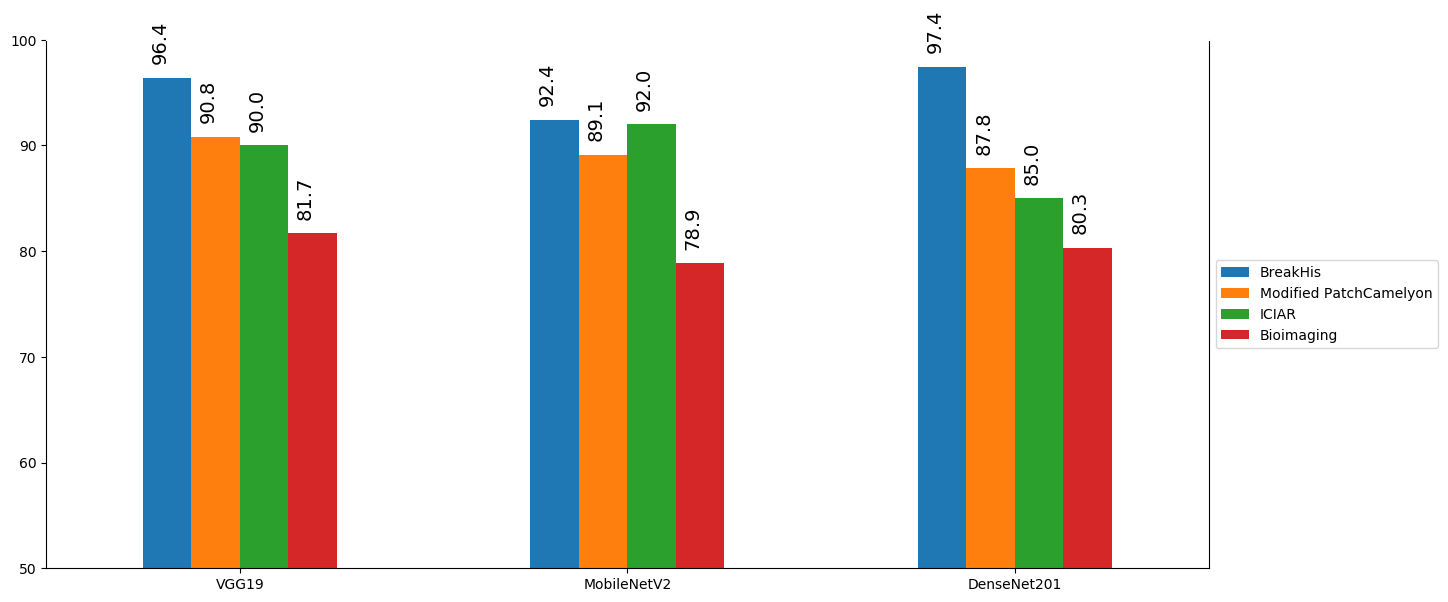}
  \caption{Classification accuracy of single classifiers of VGG19, MobileNetV2, DenseNet201}
   \label{fig:barchatSingleClassifiers}
\end{figure}

    On the other hand, the results also demonstrate that the detection rate is worst on the Bioimaging dataset with 83.10\% accuracy, 92.60\% precision, 71.42\% recall and 80.64\% F-score. Table~\ref{tab:singleClassifiers} and Figure~\ref{fig:barchatSingleClassifiers} presents the performance of the single classifiers on the four datasets. Analyzing Table~\ref{tab:singleClassifiers} and Figure~\ref{fig:barchatSingleClassifiers}, we obtain the maximum 97.42\%, 96.41\% and 92.40\% accuracies are produced on the BreakHis dataset by DenseNet201, VGG19 and MobileNetV2 models, respectively. 
    
\begin{table}[h]
    \caption{Results of accuracies obtained by single classifiers on four open access datasets.}
    \label{tab:singleClassifiers}
    \begin{tabular}{llll}
    \toprule
                   & VGG19   & MobileNetV2 & DenseNet201 \\
    \midrule
    BreakHis       & 96.41\% & 92.40\%     & 97.42\%     \\
    PatchCamelyon* & 90.84\% & 89.09\%     & 87.84\%     \\
    ICIAR          & 90.00\% & 92.00\%     & 85.00\%     \\
    Bioimaging     & 81.69\% & 78.87\%     & 80.28\%    \\    
    \bottomrule
\end{tabular}
\end{table}

\begin{table}[h]
\small
    \caption{Classification results of different state-of-the-art CNN classifiers on four datasets.}
    \label{tab:OtherCNN}
\begin{tabular}{@{}lllll@{}}
\toprule
                  & \textbf{BreakHis} & \textbf{PCamelyon*} & \textbf{ICIAR} & \textbf{Bioimaging} \\ \midrule
InceptionV3       & 87.66\%           & 87.52\%                 & 83.00\%        & 85.00\%             \\
Xception          & 86.37\%           & 88.05\%                 & 83.00\%        & 78.77\%             \\
ResNet50          & 79.48\%           & 79.06\%                  & 80.00\%        & 63.38\%             \\
InceptionResNetV2 & 92.40\%           & 89.93\%                  & 89.00\%        & 76.06\%             \\
VGG16             & 93.54\%           & 88.39\%                 & 89.00\%        & 83.10\%             \\ \bottomrule
\end{tabular}
\end{table}
\begin{table}[h]
    \caption{Comparative analysis with presented methods in the literature.}
    \label{tab:comparativeAnalysis}
\begin{tabular}{@{}lll@{}}
\toprule
\textbf{Method}          & \textbf{Dataset} & \textbf{Accuracy} \\ \midrule
Roy et al.~\cite{ROY201990}       & ICIAR            & 92.50\%           \\
Vo et al.~\cite{VO2019123}       & BreakHis         & 96.30\%           \\
Pratiher et al.~\cite{pratiher2019diving} & BreakHis         & 98.70\%           \\
Spanhol et al.~\cite{spanhol2016breast}   & BreakHis         & 84.60\%           \\
Han et al.~\cite{han2017breast}     & BreakHis         & 96.90\%           \\
Gandomkar et al.~\cite{GANDOMKAR201814}   & BreakHis         & 97.90\%           \\
Brancati et al.~\cite{brancati2018multi}  & Bioimaging       & 88.90\%           \\
Arujo et al.~\cite{araujo2017classification}     & Bioimaging       & 83.30\%           \\
Vo et al.~\cite{VO2019123}  & Bioimaging       & 99.50\%           \\ \bottomrule
\end{tabular}
\end{table}

\begin{table}[h]
\small
    \caption{Comparison of classification accuracies obtained by different machine learning models.}
    \label{tab:MLClassifiers}
    \begin{tabular}{lllll}
     \toprule
                      & BreakHis & PatchCamelyon* & ICIAR   & Bioimaging \\
    \midrule
    Decision Tree      & 91.67\%  & 76.24\%        & 77.00\% & 71.83\%    \\
    Random Forest      & 92.10\%  & 82.54\%        & 85.00\% & 69.01\%    \\
    XGBoost            & 94.11\%  & 87.15\%        & 89.00\% & 78.87\%    \\
    AdaBoost           & 91.82\%  & 76.49\%        & 79.00\% & 63.38\%    \\
    Bagging            & 94.97\%  & 88.05\%        & 87.00\% & 81.69\%    \\   
    \bottomrule
    \end{tabular}
\end{table}

    The classification results of different well-established CNN architectures, including InceptionV3, Xception, ResNet50, InceptionResNetV2 and VGG16 are summarized in Table~\ref{tab:OtherCNN}. Analyzing Table~\ref{tab:OtherCNN}, we observe that there is a level of variation in all results of datasets.  As the results confirms the proposed architecture and each of the selected single classifiers delivered higher accuracy in all of the datasets except InceptionV3 architecture for Bioimaging dataset. In Bioimaging dataset, the inceptionV3 network obtained 85.00\% accuracy which is 1.9\% lower than result obtained by proposed architecture with 83.10\% accuracy.
    
    For the sake of comparison, the performance of the proposed ensemble model is compared with the results of the previously published work for binary classification of breast cancer in Table~\ref{tab:comparativeAnalysis}. Referring to Table~\ref{tab:comparativeAnalysis}, on the BreakHis dataset, our proposed approach (98.13\% accuracy) achieved a better performance compared to the methods in~\cite{spanhol2016breast,VO2019123, han2017breast} with accuracies of 86.6\%, 96.3\% and 96.9\%, respectively. However, the result reported in the study of~\cite{pratiher2019diving} with accuracy of 98.7\% achieved better performance than our proposed method with 98.13\% accuracy with a gap of accuracy of 0.57\%. On the binary classification of ICIAR dataset, the study in~\cite{ROY201990} achieved 92.5\% while proposed method achieved 95\%. On the binary classification of Bioimaging dataset, the proposed model obtained poor results in compare with studies of~\cite{VO2019123,brancati2018multi} and only outperformed study in~\cite{araujo2017classification} [Arujo], which is slightly higher performance with a gap of accuracy of 0.7\%. Finally, for PatchCamelyon* dataset, no study reported in the literature yet.

    To validate the performance of the proposed model, we also compare the proposed method with five machine learning models, namely, Decision Tree, Random Forest, XGBoost, AdaBoost and Bagging Classifier. Table~\ref{tab:MLClassifiers} summarizes the comparison of the performance of the state-of-the-art machine learning algorithms, i.e., Decision Tree, Random Forest, XGBoost, AdaBoost and Bagging Classifier. As given in this table, the topmost result was obtained by bagging classifier with 94.97\% accuracy for BreakHis dataset. Random Forest produced 69.01\% accuracy for Bioimaging dataset, which is the worst accuracy achieved in the classification of benign and malignant cases. 

    Our proposed model in the ICIAR dataset achieved 95.00\% overall accuracy, which is the highest result reported in the literature for binary classification of this dataset with a gap in the accuracy of 5.00\% for VGG19, 3.00\% for mobileNetV2 and 10.00\% for DenseNet201. The proposed model, on the same dataset, also outperforms other machine learning models by 18.00\% for Decision Tree, 10.00\% for Random Forest, 6.00\% XGBoost, 16.00\% for AdaBoost and finally 8.00\% for Bagging Classifier. The largest gap is observed for Bioimaging dataset between the proposed model and Adaboost classifier, where the difference is more than 19.00\%. The second most significant gap is achieved for the modified PatchCamelyon dataset between the proposed model and Decision Tree classifier, where the difference is 18.40\%. The smallest gap is seen for BreakHis dataset between the proposed model and DenseNet201 architecture, where the difference is less than 1.00\%. Similar conclusions can be drawn for other models. The experiment results indicate that the performance of the proposed ensemble method yields satisfactory results and outperforms both the state-of-the-art CNNs and machine learning algorithms in cancer classification on four publicly available benchmark datasets with a large gap in terms of accuracy. The proposed method is generic as it does not need handcrafted features and can be easily adapted to different detection tasks, requiring minimal pre-processing. These datasets were collected across multiple sources with different shape, textures and morphological characteristics. The transfer learning strategy has successfully transferred knowledge from the source to the target domain despite the limited dataset size of ICIAR and Bioimaging databases. During the proposed approach, we observed that no over-fitting occurs to impact the classification accuracy adversely.
    
    The performance of all of the single classifier and the proposed ensemble model was poor on Bioimaging dataset. For this dataset, benign cases are confused with malignant cases since the morphology of some benign classes is more similar to malignant samples. Intuitively, the main reason is that the size of the Bioimaging dataset is not large enough for deep learning models to capture high-level features and distinguish classes from each other. Although, data augmentation strategies are employed to tackle this problem, but it will be more appropriate to collect more training data by increasing the number of samples rather than artificially increase the size of the dataset by data augmentation methods. Also, employing pre-trained models requires input images to be resized to a certain dimension which may discard discriminating information from this dataset.
    
\section{Conclusion}
    This paper presents an ensemble-based deep learning approach for aided diagnosis of breast cancer detection. Three well-established CNNs architectures, namely VGG19, MobileNetV2 and DenseNet201 are ensembled for feature representation and extraction using different components. The combination of such various features leads to a better generalization performance than single classifiers as counterparts. The experimental results showed that the proposed model not only outperformed the individual CNN classifiers but also outperformed state-of-the-art machine learning algorithms in all the test sets of the provided datasets. The highest and lowest performances were obtained for BreakHis and Bioimaging datasets, respectively. Thus, the deep learning-based multi-model ensemble method can make full use of the local and global features at different levels and improve the prediction performance of the base architectures across different datasets. This research is a foundation for our future publication in the integration of deep learning and blockchain technology.

\bibliographystyle{ACM-Reference-Format}
\bibliography{ref}

\end{document}